\documentclass[aps,pre,twocolumn,preprintnumbers,showpacs,superscriptaddress]{revtex4}

\usepackage{graphicx}
\usepackage{amsfonts}

\begin{document}

\title{Damage Spreading in Spatial and Small-world Random Boolean Networks}

\author{Qiming Lu}
\email{luq2@rpi.edu}
\affiliation{Department of Physics, Applied Physics, and Astronomy,
Rensselaer Polytechnic Institute, 110 8th Street, Troy, NY 12180-3590, USA}
\affiliation{Los Alamos National Laboratory, CCS-3, MS B256, Los Alamos, New Mexico 87545, USA}

\author{Christof Teuscher}
\email{christof@teuscher-lab.com}
\affiliation{Department of Electrical and Computer Engineering (ECE), Portland State University, P.O. Box 751, Portland, OR 97207-0751}

\begin{abstract}
  The study of the response of complex dynamical social, biological,
  or technological networks to external perturbations has numerous
  applications. Random Boolean Networks (RBNs) are commonly used a
  simple generic model for certain dynamics of complex
  systems. Traditionally, RBNs are interconnected randomly and without
  considering any spatial extension and arrangement of the links and
  nodes. However, most real-world networks are spatially extended and
  arranged with regular, power-law, small-world, or other non-random
  connections. Here we explore the RBN network topology between
  extreme local connections, random small-world, and pure random
  networks, and study the damage spreading with small
  perturbations. We find that spatially local connections change the
  scaling of the relevant component at very low connectivities
  ($\bar{K} \ll 1$) and that the critical connectivity of stability
  $K_s$ changes compared to random networks. At higher $\bar{K}$, this
  scaling remains unchanged. We also show that the relevant component
  of spatially local networks scales with a power-law as the system
  size N increases, but with a different exponent for local and
  small-world networks. The scaling behaviors are obtained by
  finite-size scaling. We further investigate the wiring cost of the
  networks. From an engineering perspective, our new findings provide
  the key design trade-offs between damage spreading (robustness), the
  network's wiring cost, and the network's communication
  characteristics.
\end{abstract}


\date{\today}

\maketitle

\section{Introduction}
The robustness against failures, the wiring cost, and the
communication characteristics are key measures of most complex,
finite-size real-world networks. For example, the electrical power
grid needs to be robust against a variety of failures, minimize the
wiring cost, and minimize the transmission losses. Similarly, the
neural circuitry in the human brain requires efficient signal
transmission and robustness against damage while being constrained in
volume.

In this letter, we use random Boolean networks (RBNs) as a simple
model to study the (1) robustness, i.e., the damage spreading, (2) the
wiring cost, and (3) the communication characteristics as a function
of different network topologies (local, small-world, random),
different connectivities $\bar{K}$, and different network sizes
$N$. More generally speaking, this allows us to answer the question of
{\em how much} and {\em what type} of interconnectivity a complex
network---in our case RBNs---needs in order to satisfy given
restrictions on the robustness against certain types of failure, the
(wiring) cost, and the (communication) efficiency. The work presented
here extends previous work by Rohlf et al. \cite{ROHLF_PRL2007} to new
network topologies, which are more biologically plausible, such as for
example small-world topologies.

RBNs were originally introduced by Kauffman as simplified models of
gene regulation networks \cite{Kauffman69,Kauffman93}. In its simplest
form, an RBN is discrete dynamical system, also called $NK$ network
(or model), composed of $N$ automata (or nodes), each of which
receives inputs from $K$ (either exact or average) randomly chosen
other automata. Each automaton is a Boolean variable with two possible
states: $\{0,1\}$, and the dynamics is such that
\begin{equation}
{\bf F}:\{0,1\}^N\mapsto \{0,1\}^N,
\label{globalmap}
\end{equation}
where ${\bf F}=(f_1,...,f_i,...,f_N)$, and each $f_i$ is represented
by a look-up table of $K_i$ inputs randomly chosen from the set of $N$
automata. Initially, $K_i$ neighbors and a look-table are assigned to
each automaton at random.
\begin{equation}
f_i:\{0,1\}^{K_i}\mapsto \{0,1\}.
\label{booleanfunction}
\end{equation}
An automaton state $ \sigma_i^t \in \{0,1\}$ is updated using its
corresponding Boolean function:
\begin{equation}
\sigma_i^{t+1} = f_i(x_{i_1}^t,x_{i_2}^t, ... ,x_{i_{K_i}}^t).
\label{update}
\end{equation}
We randomly initialize the states of the automata (initial condition
of the RBN). The automata are updated synchronously using their
corresponding Boolean functions.
\begin{equation}
{\bf \sigma}^{t+1} = {\bf F}({\bf \sigma}^{t}),
\label{map}
\end{equation}

In the thermodynamic limit, RBNs exhibit a dynamical order-disorder
transition at a sparse critical connectivity $K_c$
\cite{DerridaP86}. For a finite system size $N$, the dynamics of RBNs
converge to periodic attractors after a finite number of updates. At
$K_c$, the phase space structure in terms of attractor periods
\cite{AlbertBaraBoolper00}, the number of different attractors
\cite{SamuelsonTroein03} and the distribution of basins of attraction
\cite{Bastola98} is complex, showing many properties reminiscent of
biological networks \cite{Kauffman93}.

The study of the response of complex dynamical networks to external
perturbations, also referred to as damage, has numerous applications,
e.g., the spreading of disease through a population
\cite{Pastor2001,Newman2002}, the spreading of a computer virus on the
internet \cite{Cohen2003}, failure propagation in power grids
\cite{Sachtjen2000}, the perturbation of gene expression patterns in a
cell due to mutations \cite{RamoeKesseliYli06}, or the intermittent
stationary state in economic decision networks triggered by the
mutation of strategy from a few individual agents
\cite{Bassler2000}. Mean-field approaches, e.g., the annealed
approximation (AA) introduced by Derrida and Pomeau \cite{DerridaP86},
allow for an analytical treatment of damage spreading and exact
determination of the critical connectivity $K_c$ under various
constraints \cite{SoleLuque95,LuqueSole96}. However, these
approximations rely on the assumption that $N\to\infty$, which, for an
application to real-world problems, is often an irrelevant limit. A
number of studies \cite{KaufmanMihaljevDrossel05,Mihaljev2006} has
recently focused on the finite-size scaling of (un-)frozen and/or
relevant nodes in RBN with respect to $N$ with the goal to go beyond
the annealed approximation. Only a few studies, however, consider
finite-size scaling of damage spreading in RBNs
\cite{ROHLF_PRL2007,RamoeKesseliYli06,SamuelsonSocolar06}. Of
particular interest is the ``sparse percolation (SP) limit''
\cite{SamuelsonSocolar06}, where the initial perturbation size $d(0)$
does {\em not} scale up with the network size $N$, i.e., the relative
size of perturbations tends to zero for large $N$.  In
\cite{ROHLF_PRL2007} Rohlf et al. have identified a new characteristic
connectivity $K_s$ for RBNs, at which the average number of damaged
nodes $\bar d$, after a large number of dynamical updates, is
independent of $N$. This limit is particularly relevant to information
and damage propagation in many technological and natural networks. The
work in this letter extends these new findings and systematically
studies damage spreading in RBNs as a function of new network
topologies, namely local and small-world, different connectivities
$\bar{K}$, and different network sizes $N$.

\section{Damage Spreading}
For our purpose, we measure the expected damage $\bar{d}$ as the
Hamming distance between two different initial system configurations
after a large number of $T$ system updates. The randomly chosen
initial conditions differ by one bit, i.e., the damage size is $1$. As
introduced in \cite{ROHLF_PRL2007}, let $\mathcal{N}$ be a randomly
sampled set (ensemble) of $z_N$ networks with average degree
$\bar{K}$, $\mathcal{I}_n$ a set of $z_I$ random initial conditions
tested on network $n$, and $\mathcal{I^{\prime}}_n$ a set of $z_I$
random initial conditions differing in one randomly chosen bit from
these initial conditions.  Then we have
\begin{equation}
  \bar{d} = \frac{1}{z_N\,z_I}\sum_{\stackrel{n=1}{\mathcal{N}_n \in \mathcal{N}}}^{z_N}\sum_
  {\stackrel{i=1}{\vec{\sigma}_i \in \mathcal{I}_n, \vec{\sigma}^{\prime}_i \in \mathcal{I}^{\prime}_n} }^{z_I} d_i^n(T),\label{d_aver_eq}
\end{equation}
where $d_i^n(T)$ is the measured Hamming distance after $T$ system
updates. Rohlf et al. \cite{ROHLF_PRL2007} have shown that there
exists a characteristic connectivity $K_s$, at which the average
number of damaged nodes $\bar d$, after a large number of dynamical
updates, is independent of $N$.

In a given network, the nodes can be classified according to their
response to the network dynamics (e.g., see
\cite{KaufmanMihaljevDrossel05,Mihaljev2006}). This classification
allows to better explain the global network behavior with respect to
external perturbations.

A set of nodes is said to be part of the {\it frozen component} (or
{\it frozen core}) if each node's output is constant regardless of its
inputs. The states of these nodes remain constant on every attractor,
so that external perturbations cannot spread into the frozen
component. The frozen core does thus not contribute to the damage
spreading. The {\it irrelevant nodes} (or {\it irrelevant component})
are the nodes whose outputs may change, but their outputs are only
connected to either frozen or other irrelevant nodes. Again, these
nodes do not participate in the damage spreading. The remaining set of
nodes are the {\it relevant nodes} (or {\it relevant
  component}). Their state changes and each relevant node is connected
to at least another relevant node. As their name suggests, the
relevant nodes are the crucial ones, which determine the number and
the period of attractors in a given network. For our purpose, studying
the scaling behavior of the relevant component is important for the
study of damage spreading because the Hamming distance between the
damaged and the undamaged network can be viewed as a quantitative
measure of the distance between the two different attractors the
networks settle in.

In this letter, we use three exemplary types of network topologies:
(1) random, (2) spatially local, and (3) small-world. In the following
we will describe the models we used to create each of these network
topologies and what the relevant parameters are. For more details see
the text in the next three sections (\ref{sec:rand}, \ref{sec:local},
and \ref{sec:sm}).  Note that in all of these network topologies, the
links are directed, self-loops are allowed, and multiple-links between
the same pair of nodes are excluded.

\paragraph{Random topology.}
Each of the $N$ nodes has a uniform probability to be connected to any
other node in the network. The average connectivity is $\bar{K}$. This
topology corresponds to the original $NK$ model proposed by Kauffman
\cite{Kauffman69}.

\paragraph{Spatially local topology.}
$N$ nodes are uniformly and randomly distributed in a unit
$d$-dimensional spatial area (non-periodical). Each node randomly
connects to its nearest neighbors (including itself) until the
designated $\bar{K}$ is reached. This network topology can be
classified as a spatial graph. In the limit of small $\bar{K}$
($\bar{K} \ll N$), such a $d$-dimensional, spatially local network has
an average path length of $\sim N^{1/d}$ \cite{Watts03}, which is
similar to a $d$-dimensional regular lattice \cite{Newman_2003}.

\paragraph{Small-world topology.}
Starting from $2D$ spatially local networks as described above, we
apply a rewiring method to obtain a small-world network topology. The
source of every existing link will be rewired with probability $p$ to
a randomly chosen node in the network. Thus, when $p \rightarrow 0$,
we obtain the original spatially local network, while for $p
\rightarrow 1$ we obtain a random graph as described above.

\subsection{RBNs with a Random Network Topology}
\label{sec:rand}
Rohlf {\em et. al.} \cite{ROHLF_PRL2007} have systematically
investigated damage spreading, i.e., the evolution of the Hamming
distance $\bar{d}_R$, of random Boolean networks at the sparse
percolation (SP) limit. By using finite-size scaling, they found a new
characteristic connectivity $K_S = 1.875$ at which the damage
spreading is independent of the system size $N$.

In the limit of a small average degree $\bar{K} \rightarrow 0$, the
initial perturbation persists only when the damage hits nodes that are
in loops of length two or that have self-connections. For a random
network topology, the probability of generating such loops scales with
$P_{loop} \sim 1/N^2$, where $N$ is the system size. Thus, the Hamming
distance is proportional to the number of simple loops, $\bar{d}_R
\sim P_{loop} \bar{K} N \sim N^{-1}$. For large $\bar{K}$, the
relevant component grows comparable with the system size, so the
initial damage now percolates through the entire network, and we have
$\bar{d}_R \sim N$. For arbitrary $\bar{K}$, the Hamming distance
$\bar{d}_R$ scales as follows \cite{ROHLF_PRL2007}:
\begin{equation}
\bar{d}_R(\bar{K}, N) \sim a(\bar{K}) N^{\gamma_R (\bar{K})}\;,
\label{dr}
\end{equation}
where $\gamma_R \rightarrow -1$ at $\bar{K} \rightarrow 0$ and
$\gamma_R \rightarrow 1$ at large $\bar{K}$. At criticality (i.e.,
$K=2$), the asymptotic dynamics are determined entirely by the
relevant component, which scales as $n_r \sim N^{1/3}$
\cite{KaufmanMihaljevDrossel05}, thus $\gamma_R(K_c) \simeq 1/3$. As
already seen above, at $\bar{K}=K_s=1.875$ we have $\gamma_R(K_s)=0$,
and the Hamming distance $\bar{d}$ is independent of the system size
$N$ \cite{ROHLF_PRL2007}. This means that at the ``critical
connectivity of stability'' $K_s$, the damage caused by initial
perturbations is confined at a finite level (i.e., the proportion of
damage goes to zero as $N \rightarrow \infty$), regardless of the
system size $N$.

\begin{figure}[tb]
\hspace*{-0.5cm}
\includegraphics[width=3.0in]{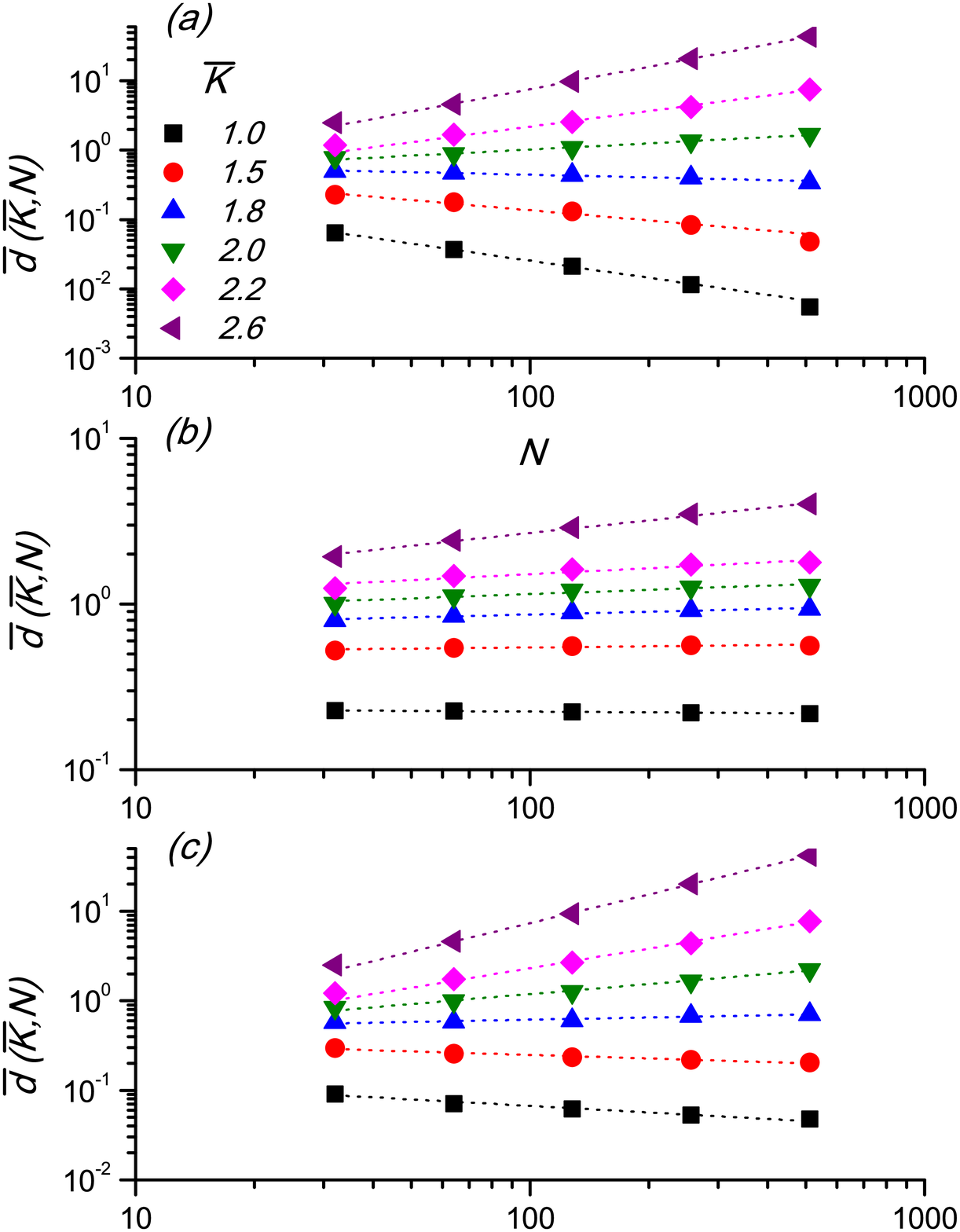}
\vspace*{-1.00truecm}
\caption{The Hamming distance $\bar{d}$ as a function of the system
  size $N$ for (a) random networks, (b) networks with spatially local
  connections, and (c) small-world networks, for different
  $\bar{K}$. $\bar{K}$ takes the values $2.6$, $2.2$, $2.0$, $1.8$,
  $1.5$, and $1.0$, from top to bottom. The data for the random
  networks confirms the data as first presented in
  \cite{ROHLF_PRL2007}. Averaged over $10,000$ randomly generated
  networks and $100$ random initial configurations for each value of
  $\bar{K}$. $T = 1000$ system updates, initial damage size $d(0)=1$.}
\label{fig.hd-n}
\end{figure}

\begin{figure}[tb]
\hspace*{-0.5cm}
\includegraphics[width=3.7in]{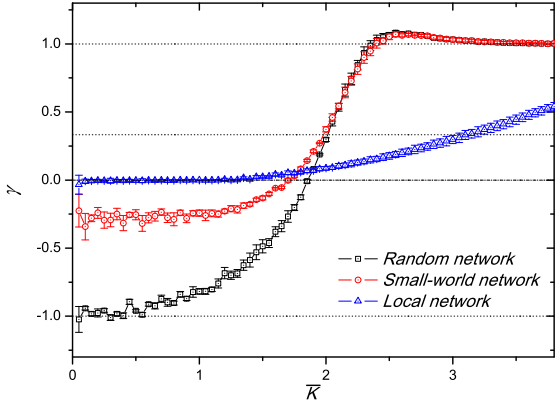}
\vspace*{-1.00truecm}
\caption{Scaling exponents $\gamma(\bar{K})$ as a function of
  $\bar{K}$ for random networks (open squares), networks with local
  connections (open triangles), and small-world networks (open
  circles). This figure is obtained from the best fit of the data of
  Fig. \ref{fig.hd-n} using Eq.~(\ref{dr}), (\ref{dl}), and
  (\ref{dsw}). The data for the random networks confirms the data as
  first presented in \cite{ROHLF_PRL2007}.}
\label{fig.gamma-k}
\end{figure}

Figure~\ref{fig.hd-n} shows the power-law dependence of the Hamming
distance $\bar{d}$ as a function of the system size $N$ for multiple
$\bar{K}$ and for our three types of random network
classes. $\gamma_R$ as a function of the average degree $\bar{K}$ is
shown in Fig.~\ref{fig.gamma-k}.

\subsection{RBNs with Spatially Local Connections}
\label{sec:local}
Many real-world networks are spatially extended and have a more
structured interconnect topology than pure random networks have. Such
networks are commonly called {\em complex networks}. Spatial networks
with local connections only, such as regular grids, have a large
average path length ($l \sim N^{1/d}$) and are highly clustered. In
this section, we look at the dynamics of spatial RBNs with local
connections only. The underlying network structure is constructed
based on the model of {\em Uniform Spatial Graphs} \cite{Watts03}, in
which vertices may connect uniformly at random to other vertices
within a spatial distance $l_c$ in $\mathbb{R}^d$.  We do this as
following: $N$ nodes are randomly distributed in a $d$-dimensional
space (only $d=2$ will be considered here), we then randomly pick a
pair of nodes $u$,$v$ and create edge $(u,v)$ if the spatial distance
is within the cut-off distance $l_c$, disallowing repeated
edges. This procedure is repeated until the required average
degree $\bar{K}$ is reached.

For a small cutoff distance, or any finite cutoff when $N \rightarrow
\infty $, e.g., $l_c \sim O(1)$, the characteristic path length
remains similar to that of $d$-dimensional regular lattices
\cite{Watts03}. For very large $\bar{K}$, the system is in the chaotic
regime and any initial damage quickly percolates through the
network. Thus, the damage is only bounded by the system size $N$,
which gives us $\bar{d}_L \sim N^1$. In the limit of $\bar{K}
\rightarrow 0$, nonzero damage can emerge only when the initial
perturbation hits a short loop of oscillating nodes. Let us assume we
have a single connection from node $A$ to node $B$ ($A \rightarrow
B$). In order to finish a simple loop between $A$ and $B$, we need to
first select node $B$ as the starting point, which has a probability
of about $\sim 1/N$. The probability to pick $A$ as a neighboring node
from $B$ to close the loop is $\sim 1/n_B$, where $n_B$ is the
possible number of $B$'s local neighbors. For a purely local network,
$n_B \ll N$. In a network of extreme local connections, the
probability of forming simple oscillating loops scales with $P_{loop}
\sim 1/n_B N \sim N^{-1}$. The number of such loops scales with $\sim
P_{loop} \bar{K} N \sim const.$, and is thus independent of the system
size $N$.  We expect to see coinciding Hamming distances at low
$\bar{K}$ for different system sizes $N$ on extremely local
networks. This remains valid until the network reaches the percolation
threshold where segregated simple loops become connected and a giant
cluster emerges. Furthermore, compared to random networks, the local
connections lowered the probability of forming the relevant component
at criticality because each relevant node needs to be controlled by
another relevant node. We thus expect that the damage increases slower
compared to random networks. In particular, the exponent is smaller
than $1/3$ at $K_c$ because $\gamma = 1/3$ at $K_c$ for random
networks \cite{ROHLF_PRL2007}.

Fig.~\ref{fig.hd-n}(a) shows the Hamming distance as a
function of the system size $N$ for different connectivities
$\bar{K}$. As one can see, for small $\bar{K}$, the damage remains
constant as $N$ increases. While for large $\bar{K}$ (above the
percolation limit) the damage spreading increases with the system size
$N$ according to a power-law. We therefore have:
\begin{equation}
\bar{d}_L(\bar{K}, N) \sim a(\bar{K}) N^{\gamma_L (\bar{K})}\;,
\label{dl}
\end{equation}
where $\gamma_L \rightarrow 0$ at $\bar{K} \rightarrow 0$, and
$\gamma_L \rightarrow 1$ for large $\bar{K}$. If we do a best fit for
the data as shown in Fig.~\ref{fig.hd-n}(a) using Eq.~(\ref{dr}),
(\ref{dl}), we obtain $\gamma_L$ as a function of $\bar{K}$. This is
shown in Fig.~\ref{fig.gamma-k}.

Finally, Fig.~\ref{fig.hd-k}(d) shows the average Hamming distance for
an initial damage size of one for local networks with different system
sizes $N$. As one can see, all curves coincide below the percolation
threshold. This confirms again our assumption of the scaling behavior
for low $\bar{K}$.

\subsection{RBNs with a Small-World Topology}
\label{sec:sm}

\begin{figure*}[t]
    \begin{center}
        \begin{minipage}[t]{0.58\linewidth}
            \hspace*{-1.00truecm}
            \raisebox{-8.2cm}{\includegraphics[width=4.7in]{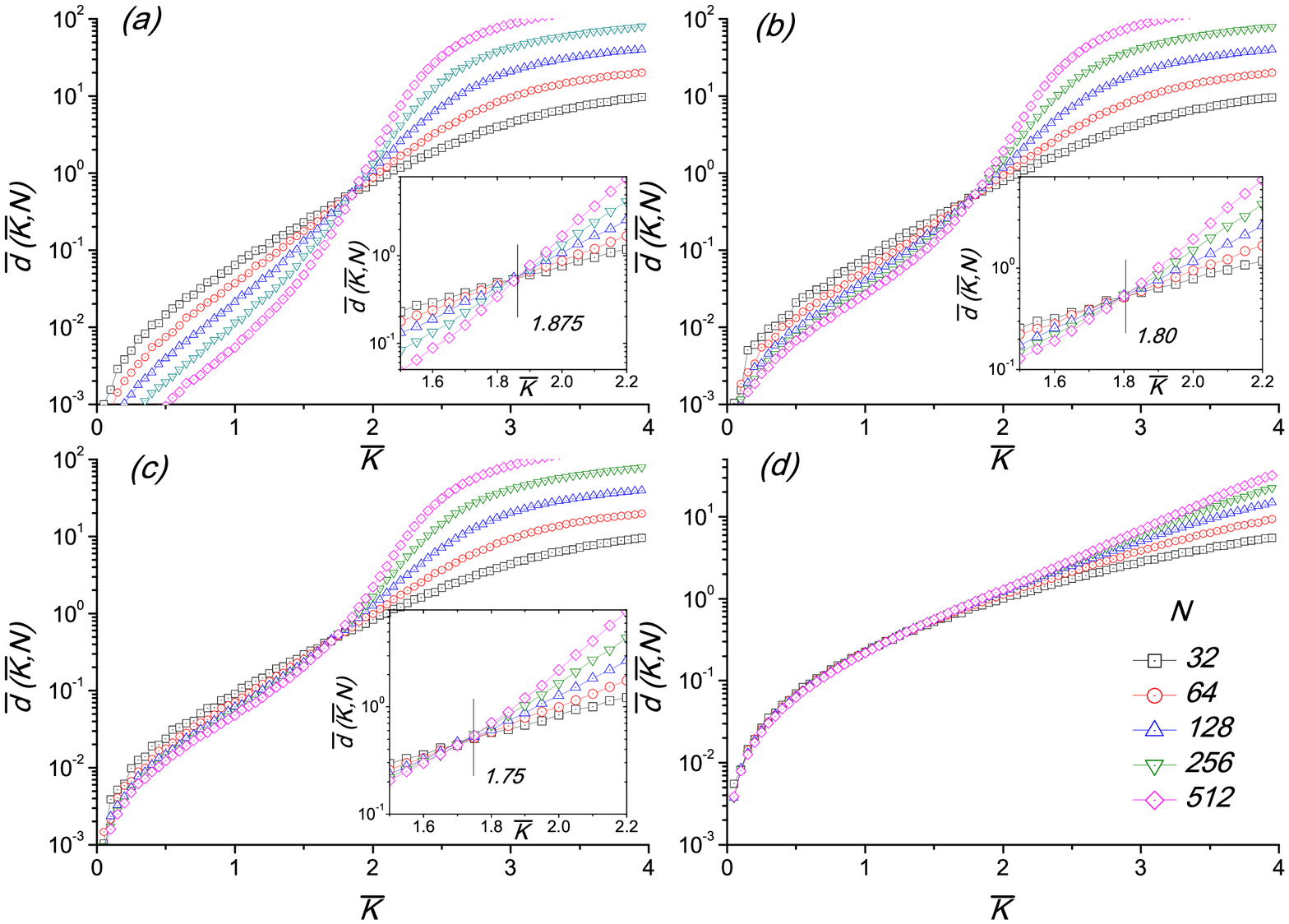}}
        \end{minipage}\hfill
        \begin{minipage}[t]{0.38\linewidth}
          \caption{Average Hamming distance (damage) $\bar{d}$ after
            $200$ system updates, averaged over $10,000$ randomly
            generated networks and $100$ random initial configurations
            for each value of $\bar{K}$. The initial damage size is
            one. Network topologies: (a) random networks ($p=1.0$,
            $q=0$), (b) small-world network with $p=0.9$ ($q=0.1$),
            (c) small-world networks with $p=0.8$ ($q=0.2$), and (d)
            networks with completely local connections ($p=0$,
            $q=1.0$). (a), (b), and (c) suggest that all curves of
            random and small-world networks for different $N$
            approximately intersect in a characteristic point $K_s$.
            $K_s$ moves toward small $\bar{K}$ as the fraction of
            local connections increases ($K_s \simeq 1.875$ in (a),
            $K_s \simeq 1.80$ in (b) and $K_s \simeq 1.75$ in
            (c)). For complete local networks all curves coincide
            below the percolation threshold independently of $N$.}
            \label{fig.hd-k}
        \end{minipage}
    \end{center}
\end{figure*}

Both purely random and purely local networks are extreme network
topologies. Many biological, technological, and social networks lie
somewhere between these two extremes and are categorized as
``small-world networks'' \cite{WATTS98}. Small-world networks
typically exhibit a number of advantages over locally connected
networks, such as a short average path length, synchronizability, and
improved robustness against certain types of failures. It is therefore
of fundamental interest to study the damage spreading in RBN networks
with a small-world interconnect topology.

Starting from the $2D$ Uniform Spatial Graph we have used above for
the locally interconnected network, we apply a simply rewiring
strategy to construct a small-world network. Each existing connection
in the Uniform Spatial Graph is rewired with probability $p$ to a
randomly chosen node. Thus, a fraction of $p$ links in the network are
random long-range links, or small-world links, while the remaining
fraction of $q=1-p$ links are local links connecting geometrically
local neighbors. We will use $q$ as the main parameter to represent
the ``strength'' of the local connections.  Combined with the system
size $N$, $Nq$ is approximately the number of nodes that have a local
connection (at the sparse percolation limit).  For $Nq \ll 1$ the
network is in the random network regime (see Sec \ref{sec:rand}); and
for $Nq \gg 1$ we obtain a spatially local network (see Sec
\ref{sec:local}).

We will now use a similar scaling approach for small-world RBNs as
presented above for local and random networks. Again, at very large
$\bar{K}$, the damage will only be bounded by the system size, thus
$\bar{d}_{SW} \sim N^1$. But for $\bar{K} \rightarrow 0$, the network
is now composed of both local and random (longer range) connections
and the probability of forming simple loops scales thus
differently. Let us assume we have a local link that has already been
connected ($A \rightarrow B$). The probability of having such a local
link is $q$, and to complete a simple loop that contains this local
connection, we first need to pick node $B$ with probability $1/N$. Node $B$ will then establish connections again with his local neighbors with
probability $q$, and finally choose node $A$ to finish the loop with
probability $1/n_B$. Thus, the final probability of having a simple loop
in this case scales with $P_{loopL} \sim q^2/N$. Similarly the
probability of generating a simple loop involving random long-range
links is $P_{loopR} \sim p^2/N^2$. We compare these two probabilities
by dividing one by another:
\begin{equation}
\frac{P_{loopL}}{P_{loopR}} = \frac{q^2}{N} / \frac{p^2}{N^2} = (Nq) \cdot \frac{q}{p^2}\;.
\label{plpr}
\end{equation}
In the spatially local network limit ($Nq \gg 1$), $P_{loopL}$ is the
leading term and the scaling follows Eq.~(\ref{dl}). In the random
network limit ($Nq \ll 1$), $P_{loopR}$ dominates and the scaling
follows Eq.~(\ref{dr}). However, when the network is in the
small-world regime, $P_{loopL}$ and $P_{loopR}$ become
comparable. With some corrections, we therefore have:
\begin{equation}
P_{loopSW} = P_{loopL} + P_{loopR} = \frac{q^2}{N} + \frac{p^2}{N^2} \simeq \frac{1}{N^\beta}\;,
\label{psw}
\end{equation}
where $\beta$ is somewhere between $1$ and $2$ and depends on the
value of $Nq$. The damage spreading scales therefore with
$\bar{d}_{SW} \sim N^{1-\beta}$. And for general $\bar{K}$, we
obtain:
\begin{equation}
\bar{d}_{SW}(\bar{K}, N) \sim a(\bar{K}) N^{\gamma_{SW} (\bar{K})}\;,
\label{dsw}
\end{equation}
where $\gamma_{SW}$ is somewhere between $1-\beta$ to
$1$. Fig.~\ref{fig.gamma-k} shows $\gamma_R$, $\gamma_L$, and
$\gamma_{SW}$. As one can see, $\gamma_{SW}$ goes from $1-\beta$,
which is below zero, to $1$ as $\bar{K}$ increases. In addition, the
critical connectivity $\bar{K}$, where $\gamma_{SW} = 0$, is different
from random networks and depends $q$. In random networks this point is
defined as the critical degree of stability $K_s$
\cite{ROHLF_PRL2007}. Our results show that the introduction of local
connections in random networks changes $K_s$ toward lower
$\bar{K}$. As we have seen above, in extreme local networks $K_s$ is
undefined because the Hamming distance for different system sizes $N$
simply coincide below the percolation threshold. Fig.~\ref{fig.hd-k}
(c) shows the deviation of $K_s$ from the observed value $K_s = 1.875$
for random networks.

\subsection{Finite-size Scaling in Random Small-World RBNs}

\begin{figure}[tb]
\hspace*{-0.5cm}
\includegraphics[width=3.7in]{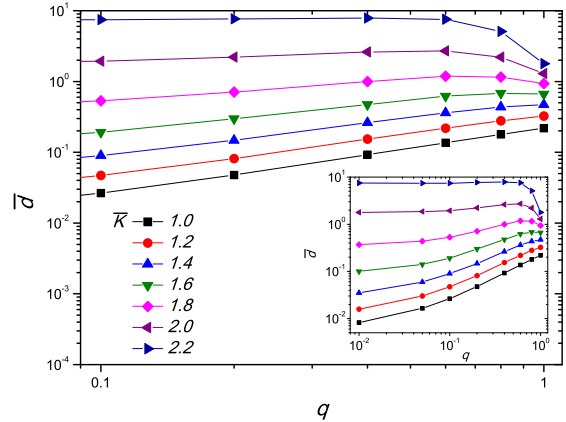}
\vspace*{-1.00truecm}
\caption{Hamming distance $\bar{d}$ for different average degrees
  $\bar{K}$ as a function of the density of local connections $q$.
  The inset shows $\bar{d}$ in the range of random networks ($q
  \rightarrow 0$) to local networks ($q \rightarrow 1$). The different
  curves range from (top to bottom) $\bar{K}=2.2$ to $\bar{K}=1.02$
  with an interval of $0.2$. }
\label{fig.hd-q}
\end{figure}

\begin{figure}[tb]
\hspace*{-0.5cm}
\includegraphics[width=3.7in]{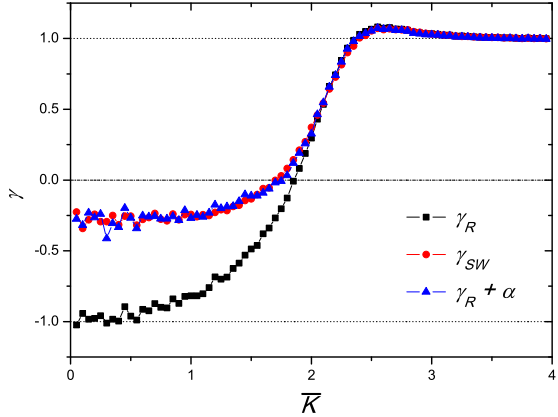}
\vspace*{-1.00truecm}
\caption{Reconstructed $\gamma_{SW}$ from Eq.~\ref{gamma-r-sw} by
  measuring $\alpha(\bar{K})$ from best fits of Fig.~\ref{fig.hd-q} at
  $q=0.2$ ($p=0.8$). Squares (black) and circles (red) are measured
  $\gamma_R$ and $\gamma_{SW}$, respectively, from
  Fig.~\ref{fig.hd-n}, while triangles (blue) are reconstructed
  $\gamma_{SW}$ from $\gamma_R$ and $\alpha(\bar{K})$.}
\label{fig.gamma-recon}
\end{figure}

\begin{figure*}[tb]
\hspace*{-0.5cm}
\includegraphics[width=7.8in]{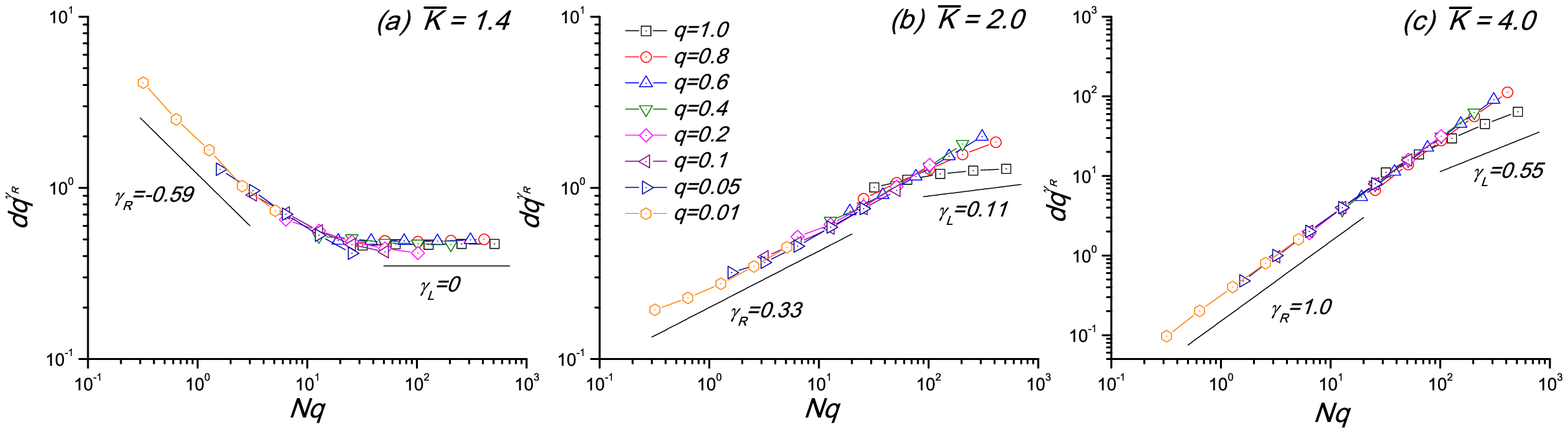}
\vspace*{-0.50truecm}
\caption{Scaling plots of the Hamming distance in small-world networks
  as predicted by the finite-size scaling argument
  (Eq.~(\ref{dg})). (a) $\bar{K}=1.4$; (b) $\bar{K}=2.0$; and (c)
  $\bar{K}=3.0$. The straight line segments correspond to asymptotic
  power-law behavior of the scaling function $g(x)$ with exponents
  $\gamma_R$ and $\gamma_L$, measured from Fig.~\ref{fig.gamma-k} at
  given $\bar{K}$, for small and large arguments, respectively, as
  described in the text (Eq.~(\ref{gx})).}
\label{fig.g-x}
\end{figure*}

While Eqs.~(\ref{dr}) and (\ref{dl}) provide the scaling of the
Hamming distance for random and local networks, we are interested in
this section how $q$ (i.e., the fraction of local links) affects the
damage spreading.

Fig.~\ref{fig.hd-q} shows the hamming distance as a function of
$q$. For sufficiently large $Nq$, i.e., close to the local network
limit, we assume (see Fig.~\ref{fig.hd-q}) that the Hamming distance
approaches an asymptotic power law $\bar{d}_{SW} \sim
q^{\alpha(\bar{K})}$. On the other hand, the Hamming distance also
depends on the system size with $\bar{d}_{SW} \sim N^{\gamma_{SW}
  (\bar{K})}$ (see Eq.~(\ref{dsw})). Thus, in the small-world regime
(close to the local network limit) the Hamming distance depends on
both the system size and the density of local (random) connections:
\begin{equation}
\bar{d}_{SW}(q, \bar{K}, N) \sim q^{\alpha(\bar{K})} N^{\gamma_{SW} (\bar{K})}\;.
\label{dqsw}
\end{equation}
When $q \rightarrow 1$, $\gamma_{SW}(\bar{K}) \rightarrow
\gamma_L(\bar{K})$, so $\bar{d}_{SW} \rightarrow \bar{d}_L$. While in
the random network limit ($Nq \ll 1$), $\bar{d}$ only depends on $N$,
$\bar{d}_R \sim N^{\gamma_R(\bar{K})}$ with $\gamma_R(\bar{K})$ as
illustrated in Fig.~\ref{fig.gamma-k}. To connect the above two cases
and to capture the finite-size behavior in the small-world regime, one
can construct the full scaling behavior of $\bar{d}_{SW} (q, \bar{K},
N)$:
\begin{equation}
\bar{d}_{SW}(q, \bar{K}, N) \sim q^{\alpha(\bar{K})} N^{\gamma_{SW} (\bar{K})} f(Nq)\;,
\label{df}
\end{equation}
where $f(x)$ is a scaling function such that
\begin{equation}
f(x)\sim\left\{
    \begin{array}{ll}
        x^{-\alpha} & \mbox{if $x$$\ll$$1$} \\
        {\rm const.} & \mbox{if $x$$\gg$$1$}
    \end{array}
\right. \;.
\end{equation}
The random network limit is obtained provided that
\begin{equation}
\bar{d} \sim q^{\alpha(\bar{K})} N^{\gamma_{SW} (\bar{K})} (Nq)^{-\alpha} \sim N^{\gamma_{SW}-\alpha} \sim N^{\gamma_R}\;,
\label{}
\end{equation}
i.e.,
\begin{equation}
\gamma_R = \gamma_{SW} - \alpha(\bar{K})\;.
\label{gamma-r-sw}
\end{equation}
Given $\gamma_R$ we can express $\gamma_{SW}$ by measuring
$\alpha(\bar{K})$ at different $\bar{K}$. Fig.~\ref{fig.gamma-recon}
shows the reconstructed $\gamma_{SW}$ with the measured data, which
satisfy the above proposed asymptotic scaling relation.

To analyze our data, Eq.~(\ref{df}) can also be written as
\begin{equation}
\bar{d}_{SW}(q, \bar{K}, N) \sim \frac{(Nq)^{\gamma_{SW}}} {q^{\gamma_{SW}-\alpha}} f(Nq) \sim \frac{1}{q^{\gamma_R}} g(Nq)\;,
\label{dg}
\end{equation}
where $g(x)=x^{\gamma_{SW}}f(x)$. Thus plotting $\bar{d}q^{\gamma_R}$
vs. $Nq$ should yield coinciding data with $g(x)$. The limits of random
and spatially local networks correspond to the asymptotic small and
large argument of $g(x)$, which gives us the exponents $\gamma_R$, and
$\gamma_L$,
\begin{equation}
g(x)\sim\left\{
    \begin{array}{ll}
        x^{\gamma_{\rm R}} & \mbox{if $x$$\ll$$1$} \\
        x^{\gamma_{\rm SW}} \rightarrow x^{\gamma_{\rm L}} & \mbox{if $x$$\gg$$1$}
    \end{array}
\right. \;.
\label{gx}
\end{equation}

Fig.~\ref{fig.g-x} shows the scaling plots of the Hamming distance as
a function of the product of the system size $N$ and strength of the
local connections $q$, as predicted by the proposed finite-size
scaling for small-world RBNs. Also, given that $\gamma_R$ and
$\gamma_L$ are functions of the average degree $\bar{K}$, as shown in
Fig.~\ref{fig.gamma-k}, the shapes of $f(x)$ or $g(x)$ also changes
with $\bar{K}$. As one can see in Fig.~\ref{fig.g-x}(a)-(c), $g(x)$
coincides under different $\bar{K}$. In addition, the asymptotic
behavior of $g(x)$ at $x \ll 1$ and $x \gg 1$ agree very well with our
measured ``phenomenological'' exponents $\gamma_R(\bar{K})$, and
$\gamma_L(\bar{K})$ at $\bar{K}=1.4$, $\bar{K}=2.0$, and
$\bar{K}=4.0$, respectively.

\section{Wiring Cost}

\begin{figure}[tb]
\hspace*{-0.5cm}
\includegraphics[width=3.7in]{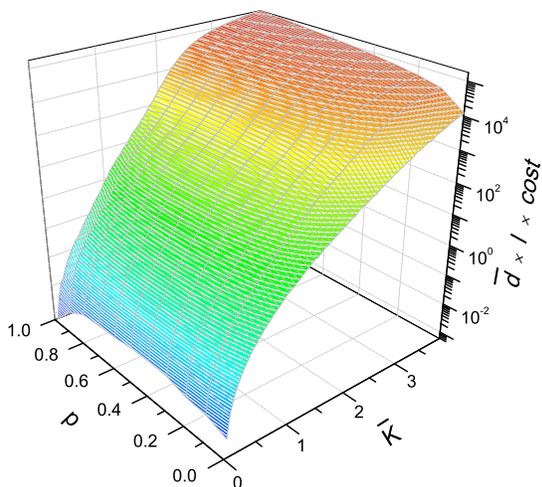}
\vspace*{-1.00truecm}
\caption{The product of damage size $d$, the network wiring cost
  $cost$, and the average shortest path length $l$ as a function of
  the average connectivity $\bar{K}$ and the density of random
  connections $p$. The color density corresponds to the value of $d
  \times l \times cost$}
\label{fig.cost-m}
\end{figure}

\begin{figure}[tb]
\hspace*{-0.5cm}
\includegraphics[width=3.7in]{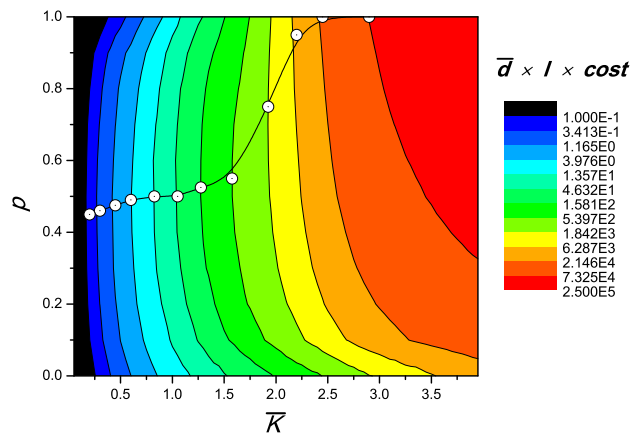}
\vspace*{-1.00truecm}
\caption{Contour projection of Fig.~\ref{fig.cost-m}. The color
  density corresponds to the value of $d \times l \times cost$. At
  $p=0$ the network topology is completely local; $p=1$ corresponds to
  the random network; and for $0<p<1$ the network is in the
  small-world regime. The circles indicate the position of lowest
  possible $\bar{K}$ and corresponding $p$ for given tolerance level
  of $d \times l \times cost$.}
\label{fig.cost-c}
\end{figure}

From an engineering perspective, one wants to typically minimize the
wiring cost of a network, maximize the communication characteristics,
and maximize the robustness against failures. The electric power grid
is a good example and so are nano-scale interconnect networks
\cite{teuscher07:chaos}. In this section we will look at these three
trade-offs for RBNs.

The average shortest path length is generally a good measure for the
communication characteristics of a complex network. In a directed
network we define $l$ as the mean geodesic (i.e., shortest) distance
between vertex pairs in a network \cite{Newman_2003}:
\begin{equation}
l = \frac{1}{n(n-1)}\sum_{i \neq j}d_{ij}\;,
\label{path}
\end{equation}
where $d_{ij}$ is the geodesic distance from vertex $i$ to vertex
$j$. Here we have excluded the distance from each node to
itself. Eq.~\ref{path} will be problematic if the network has more
than one component, which is very likely for small $\bar{K}$. To avoid
the problem of disconnected networks, we compute the average path
length only for those vertex pairs that actually have a connecting
path between them.

For real-world networks, if shortcuts, i.e., the random small-world
links, have to be realized physically, the cost of a long-range
connection is likely to grow with its length. E.g., the power
consumption for wireless broadcast communication in free space
generally could be a cubic power of geometric distance, while the
implementation of directional antenna will reduce the transmission
cost significantly \cite{Huang06}. Petermann et al.
\cite{Petermann05} discuss the wiring-cost for some spatial
small-world networks ranging from integrated circuits, the Internet to
cortical networks. For simplicity, we assume here that the wiring cost
has a linear dependency on the geometrical distance between two nodes.

Fig.~\ref{fig.cost-m} shows the product of the wiring cost $cost$, the
average shortest path length $l$, and the damage size $\bar{d}$ as a
function of the average connectivity $\bar{K}$ and the density of
random connections $p$.  In addition, Fig.~\ref{fig.cost-c} shows the
contour projection of Fig.~\ref{fig.cost-m} using the same data. For
example, it shows that given a specific target average connectivity
$\bar{K}$, networks with a high density of random connections (i.e.,
close to random networks, $p \to 1$) will have a similar overall
performance, cost, and robustness as networks that have more local
connections ($p \to 0$), however, whereas $p \to 1$ networks have a
high wiring cost, a low average path length, and a low damage
resistivity, $p \to 0$ networks have a low wiring cost, a high average
path length, and a high damage resistivity.  Fig.~\ref{fig.cost-c}
also tells us what will be the lowest $\bar{K}$ and corresponding $p$
allowed at a given robustness level. This is indicated by the circles.

\section{Conclusion}
We have systematically investigated the damage spreading in spatial
and small-world random Boolean networks. We have found that (1)
spatially local connections change the scaling of the relevant
component at very low connectivities ($\bar{K} \ll 1$) and (2) that
the critical connectivity of stability $K_s$ changes compared to
random networks \cite{ROHLF_PRL2007}. At higher $\bar{K}$, this
scaling remains unchanged. We also show that the relevant component of
spatially local networks scales with a power-law as the system size N
increases, but with a different exponent for local and small-world
networks. The scaling behaviors are obtained by finite-size
scaling. In addition, we have investigated the trade-offs between the
wiring cost of the networks, the communication characteristics, and
the robustness, i.e., the damage spreading. From an engineering
perspective, one typically wants to minimize the wiring cost, maximize
the communication characteristics, e.g., the shortest path between any
two nodes, and maximize the robustness against failures. Our new
findings provide these key trade-offs and allow to determine the
lowest connectivity $\bar{K}$ and the amount of randomness $p$ in a
network for a given robustness, average path length, and wiring cost.

Future work will focus on the investigation of real-world networks and
the application of our methodology to make them more robust, cheaper,
and more efficient.

\begin{acknowledgments}
We gratefully acknowledge the support of the U.S. Department of Energy
through the LANL/LDRD Program for this work. The authors thank Natali
Gulbahce, Gyorgy Korniss, and Thimo Rohlf for the helpful comments on
this work.
\end{acknowledgments}

\end{document}